\documentclass[pre,superscriptaddress,twocolumn,sort&compress,round,footinbib]{revtex4-1}

\usepackage{graphicx}
\usepackage{gensymb}
\usepackage{epstopdf,epsfig}
\usepackage{natbib}
\usepackage[dvipsnames]{xcolor}
\usepackage{hyperref}
\hypersetup{
    colorlinks = true,
    urlcolor   = blue,
    citecolor  = black,
}
\newcommand{\SImum}{\textrm{\textmu{}m}}
\definecolor{firebrick}{rgb}{0.7, 0.13, 0.13}

\begin{document}

\title[Numerical approach to clogging]{Clogging of non-cohesive suspensions through constrictions using an efficient unresolved CFD-DEM solver}

\author{Edgar Ortega-Roano}
\affiliation{Physics of Fluids, University of Twente, The Netherlands}

\author{Mathieu Souzy}
\affiliation{INRAE, Aix-Marseille Univ, UMR RECOVER, 13182 Aix-en-Provence, France}
\thanks{EOR and MS have contributed equally to this work}


\author{Thomas Weinhart}
\affiliation{Multi Scale Mechanisms, University of Twente, The Netherlands}

\author{Devaraj van der Meer}
\affiliation{Physics of Fluids, University of Twente, The Netherlands}

\author{Alvaro Marin}
\email{a.marin@utwente.nl}
\affiliation{Physics of Fluids, University of Twente, The Netherlands}

\begin{abstract}

When objects are forced to flow through constrictions their transport can be frustrated temporarily or permanently due to the formation of arches in the region of the bottleneck. While such systems have been intensively studied in the case of solid particles in a gas phase being forced by gravitational forces, the case of solid particles suspended in a liquid phase, forced by the liquid itself, has received much less attention. In this case,  the influence of the liquid flow on the transport efficiency is not well understood yet, leading to several apparently trivial, but yet unanswered questions, e.g., would an increase of the liquid flow improve the transport of particles or worsen it?
Although some experimental data is already available, it lacks enough detail to give a complete answer to such a question. Numerical models would be needed to scrutinize the system deeper. 
In this paper, we study this system making use of an advanced discrete particle solver (MercuryDPM) and an approximated numerical model for the liquid drag and compare the results with experimental data. 

\end{abstract}
\maketitle

\section{Introduction} 

Many-body systems flowing through constrictions, like sand in an hourglass, might flow continuously, but they might also \textit{tick} \cite{wu1993hour}, i.e., flow intermittently, or even get permanently clogged. The case of an hourglass is quite paradigmatic since time immemorial and it has been studied more intensively in the last decades. We know that, if the bottleneck is wide enough for particles to flow continuously, the particle transport rate depends on the relative size of the particle to the bottleneck with a power $N-1/2$ (where $N=2,$ 3 is the dimensionality of the system)\cite{beverloo1961flow}. However, this continuous flow is compromised when the size of the particles falls in the same order of magnitude of the bottleneck size. 

Most constricted flows display qualitatively similar behaviour regardless of their nature (grains, suspensions, pedestrians or animals). For example, when the flow of bodies becomes intermittent, the probability distribution of time lapses between the passages of consecutive bodies present remarkable power-law tails in all cases, with an  exponent that depends on the flow regime \cite{Zuriguel2014}.


The statistical framework sketched above holds as long as the passing bodies remain as individual entities. This condition does not hold, for example, for particles of colloidal size, which typically feel a strong attraction towards solid boundaries and to other particles due to van der Waals forces. Consequently, when colloidal particle suspensions are forced through constrictions, they tend to form aggregates either by successive deposition at the constriction's walls, growing up to sizes capable of blocking the flow \cite{Delouche2021}, or even forming aggregates further upstream, large enough to sieve the constriction. The clogging mechanisms for such suspensions have been covered often in the literature and they strongly depend on the physical-chemistry of the system, which determine the strength of the particle-particle attraction \cite{Wyss2006,Agbangla2012,dersoir2015clogging,duru2015three,Delouche2020}.

On the contrary, non-cohesive suspensions typically clog constrictions purely by mechanical forces much like its dry counterparts. In previous studies \cite{Marin2018,Souzy2020,souzy2022role}, we showed that non-cohesive suspensions follow the same statistical framework as granular materials \cite{Zuriguel2014}, and therefore concluded that the clogging mechanisms must be identical. 
Interestingly, this analogy seems to hold regardless of the driving method, either by gravity \cite{Koivisto2017}, by pressure \cite{souzy2022role}, by pumps \cite{Souzy2020} and even for self-propelled suspensions \cite{Bechinger2022Activeclog}.
While silos are mostly driven simply by gravity (i.e., a body force that acts uniformly on all particles), suspensions can be driven by the drag produced by the liquid flow, which adds an additional control parameter into the system that could be potentially used to optimize the transport of material through the bottleneck. Unfortunately, that is not a trivial matter. Indeed, an increase in pressure (or in liquid volume rate) may have a detrimental (or little) effect on the transport of material through the bottleneck \cite{souzy2022role}.

Recent experiments have revealed the crucial role of the interstitial liquid flow in the clogging of non-cohesive suspensions \cite{souzy2022role}, but several important details remains elusive. Numerical simulations could make a significant contribution to answer these questions. 

For decades now, numerical discrete particle methods are able to solve the dynamics of particles exerted by forces and torques of different nature (gravitational, electromagnetic, etc.), as well as the mutual interactions among the particles. Some are even capable of including contact forces of different nature (adhesive, elastic, plastic and/or viscous). 

To solve numerically the interaction between particles and fluids, two approaches are typically employed: (1) a direct computation of the liquid flowing through the pore space, which accurately resolves the viscous drag or (2) modeling the viscous drag (drag closure) based on the space-averaged flow around each particle. These are respectively known as (1) \textit{resolved} and (2) \textit{unresolved methods}. {Resolved methods} are typically preferable when the question to answer demands accuracy and precision, but the method requires fine computational meshes to resolve the flow accurately. Consequently, computation times are high, and they are typically limited to small-sized systems. A good example of resolved methods are lattice-Boltzmann simulations \cite{Hidalgo2018,vanderhoef2005lattice}, in which both solid and fluid phase are solved through a discretized version of the Boltzmann equation. 
Unresolved methods are more convenient when averaging the flow field yields a good approximation of the final solution. For example, in CFD-DPM (computational fluid dynamics - discrete particle method), locally averaged equations for flow (CFD) and Newton's equations of motion for the discrete particle system (DPM) are solved first independently. Then, fluid-particle interactions need to be defined through a drag closure model. This approach allows for computations with millions of particles in average-sized computer clusters and has been successfully used in systems as fluidized beds \cite{Deen2007Review,Kuipers2005lattice}, granular batch sedimentation \cite{zhao2014investigation}, particle beds \cite{vaango2018beds} and solid-fluid mixing \cite{blais2016mixing}.

 
One of the main features that makes clogging of suspensions complex (but also interesting) is its stochastic nature, both for the clog formation as for the clog destruction. Consequently, long simulations are required to gain enough statistics and a fully resolved method would require a substantially high computational cost. 
Therefore, in this work we make use of an unresolved method: we numerically integrate the dynamics of a non-cohesive suspension flowing through a bottleneck with the discrete particle solver MercuryDPM, and approximate the fluid-particle interaction using well-known drag closure relations. The analogous experimental system presents high particle monodispersity and homogeneity, and therefore seems ideal for a comparison with such a numerical model.
The comparison between numerics and experiments is made by computing the statistical distributions of burst durations and clogging/arrest times for both numerics and previously published experimental data \cite{souzy2022role,Souzy2020}.

The paper is organized as follows: In section \ref{sec:methods} we describe the numerical method employed and in section \ref{sec:numericalresults}, we present the results obtained with it. Here we perform a direct comparison of the numerical and experimental results. We finalize the paper with a final conclusion, including a perspective on future research in section \ref{sec:conclusion}.

\section{Numerical Methods}\label{sec:methods} 

Simulations are performed using the open source code MercuryDPM, created to perform Discrete Particle Method (DPM) simulations \cite{weinhart2020fast}. In this case, the code is applied to simulate the motion of particles inside a micro-channel. A Poiseuille flow profile is imposed in the channel and causes a drag onto the particles, pushing them towards the constriction. This drag is corrected based on the particle packing fraction following the approach proposed by \textcite{vanderhoef2005lattice}. MercuryDPM numerically computes the forces and torques that stem either from external body forces (such as a drag force originating from the liquid), or from particle interactions (such as contact forces). Although MercuryDPM has been developed extensively for dry granular applications, it could also be adapted to include hydrodynamic interactions such as lubrication forces resulting from the thin layer of viscous fluid that separates nearly touching particles, as it will be described in the following.

The contact between particles is modelled using the so called Hertzian Spring Dashpot (HSD) model \cite{antypov2011analytical}, where the normal repulsion force between two spherical particles getting into contact is 

\begin{equation}
    F_\mathrm{Hertz} = \frac{4 E_\mathrm{eff} \sqrt{r_\mathrm{eff} } }{3} \delta^{3/2} ,
\end{equation}

where $E_\mathrm{eff} = E/2(1-\nu^2)$ is the effective Young modulus with $E$ being the Young modulus of the particle material and $\nu$ its Poisson ratio, $r_\mathrm{eff}$ is the effective radius, and $\delta$ refers to the overlap between particles: $\delta=0$ for non-touching particles, and $\delta>0$ for overlapping particles. Using such model, particles do not deform but overlap keeping their spherical shape; thus, as the overlap $\delta$ increases, the repulsion force $F_\mathrm{Hertz}$ increases to separate the particles.
Note also that for two particles $i$ and $j$ in contact, $1/r_\mathrm{eff} = 1/r_i + 1/r_j$ where $r$ is the radius of the particle. For a monodisperse suspension, it results $r_\mathrm{eff} = d/4$ where $d$ is the particle diameter.

Using the HSD model, the contact time during particle collisions is given by 

\begin{equation}
    \tau_\mathrm{Hertz} = 2.214 \left( \frac{\rho}{E_\mathrm{eff}} \right)^{2/5}
    \frac{d}{v_c^{1/5}} ,
\end{equation}

where $\rho$ is the particle density, and $v_c$ the typical collision velocity. In order to accurately resolve the contact, a typical collision velocity between particles is set as a fraction of the average flow velocity $v_c = 0.1 \bar{v}$, which yields small enough contact times to avoid any numerical errors. We have checked that a different choice for the collision velocity does not visibly affect the results, as expected due to the weak dependency of the particle collision contact time $\tau_\mathrm{Hertz}$ with $v_c$. The time step for the simulations is then set as $\tau_\mathrm{Hertz}/50$.

As stated previously, as hydrodynamic interactions we include the normal component of the lubrication force $\mathbf{F}_i^{\rm L}$ experienced by a particle $i$ due to nearby particles $j$ using the expression  
\begin{equation}
    \mathbf{F}_i^{\rm L} = \sum_{i \neq j} \frac{ 3 \pi \eta d^{2} }{8 h_{i j}}
    \mathbf{ \hat{n} } ( \mathbf{u}_j - \mathbf{u}_i ) \cdot \mathbf{ \hat{n} } ,
    \label{eq:3}
\end{equation}

where $\eta$ is the fluid viscosity and $\mathbf{u}_i$ the velocity of particle $i$ \cite{batchelor1972hydrodynamic}. If the distance between the position of the centers of particles $i$ and $j$ is $|\mathbf{x}_{ij}| = | \mathbf{x}_{i} - \mathbf{x}_j |$, then the separation distance of these particles is $h_{ij} = |\mathbf{x}_{ij}| - r_i - r_j$, and the normal unit vector pointing from particle $j$ to $i$ is thus $\mathbf{ \hat{n} } = \mathbf{x}_{ij}/|\mathbf{x}_{ij}|$. Lubrication forces are pair-wise short-ranged hydrodynamic interactions for  particles $(i,j)$ satisfying $2 \xi \leq h_{ij} \leq d/2$ (for the monodisperse case), with $\xi$ the roughness of the particle. As expected, they are attractive for particles with diverging trajectories and repulsive for converging ones. 
Given the high particle packing fractions that we are considering here, lubrication forces are the only relevant hydrodynamic forces to consider and we can safely neglect the role of longer ranged interactions.

\begin{figure}
\includegraphics[width=\columnwidth]{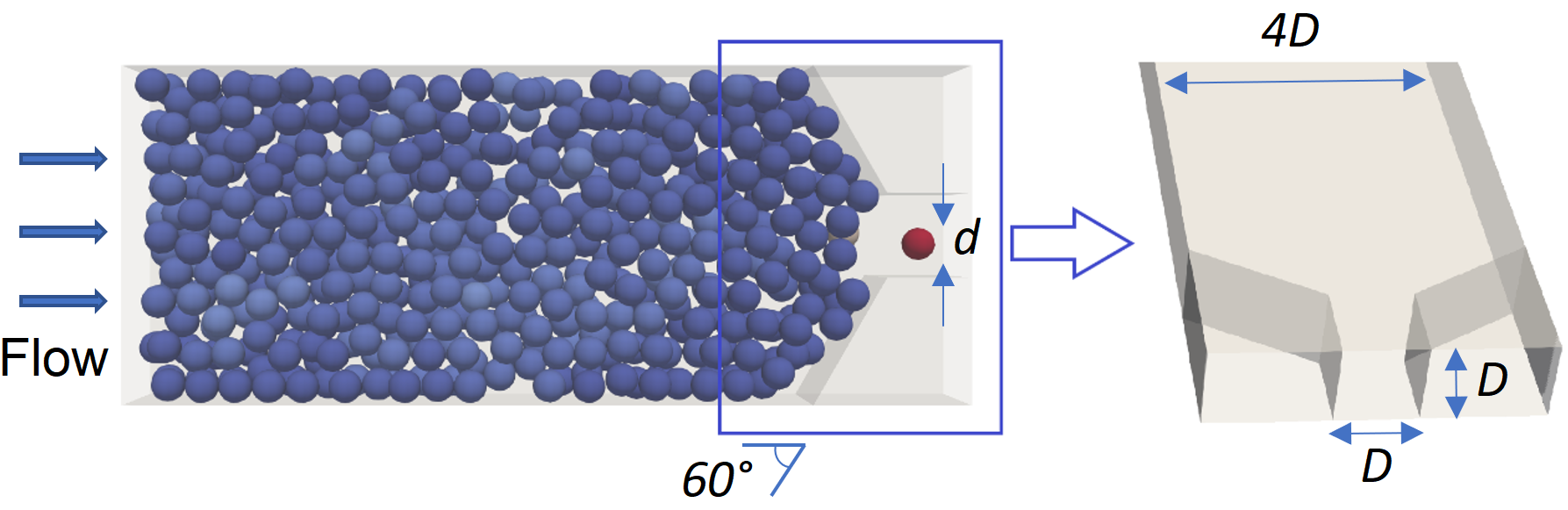}
\caption{Schematic of the constricted channel of width $4D$ used in the numerical simulations, with $D = 100 \; \mathrm{\mu m}$ the size of the constriction as shown in the right figure. The constriction angle was set at $60^{\circ}$. A suspension of particles with a diameter $d$ is forced through the constriction. A Poiseuille like flow is imposed on the left of the channel and far from the constriction.}
\label{fig:1}
\end{figure}

Regarding the geometry of the numerical setup, the suspension flows in a rectangular channel of thickness $D$ and width $4D$, which reduces to a square cross section of $D\times D$ to form the constriction, as depicted in Fig. \ref{fig:1}. This is achieved by a linear narrowing of the channel with a half angle of $60^{\circ}$. $D$ was chosen equal to $100 \; \mathrm{\mu m}$ to match the experimental setup used in the experiments \cite{Marin2018,Souzy2020,souzy2022role}. The channel length upstream the constriction is chosen to be $10D$. No significant quantitative difference was found in the results by extending the channel length from $10D$ to $15D$ and $20D$, so $10D$ was chosen for sake of faster time computation.

The flow in the channel out of the constriction is imposed by a Poiseuille flow profile \cite{bruus2007theoretical} with an average flow velocity of $\bar{v} = 5 \: \mathrm{mm/s}$, based on the typical particle velocities found in experiments \cite{Souzy2020,souzy2022role}. Close to the constriction, the fluid flow is numerically computed on a mesh using the software COMSOL. A closest mesh neighbor interpolation then provides the flow velocity for each particle position. Liquid flow rate is kept constant in the presence of particles by imposing a fluid velocity correction $V_0/\epsilon$, where $V_0$ is the computed fluid velocity in the absence of particles and $\epsilon$ accounts for the local porosity of the particle suspension. 



\begin{figure}
\includegraphics[width=\columnwidth]{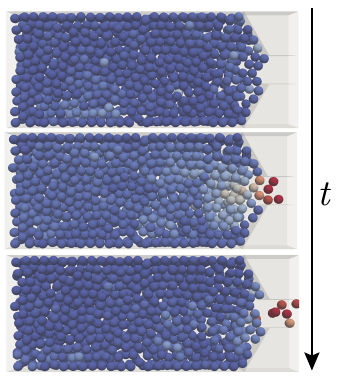}
\caption{Successive snapshots of a burst in a suspension of particles with a diameter $d = 33 \; \mathrm{\mu m}$ that intermittently flows through a constriction having a neck width and height of $D = 100 \; \mathrm{\mu m}$ with $D/d = 3.03$. From top to bottom: arch of particles formed around the neck in a clogged state, multiple particles escape after breaking of the arch, a new arch is formed at the constriction. Particles are color-coded according to their velocity: immobile particles are shown in dark blue, and fast-moving particles in dark red.}
\label{fig:2}
\end{figure}

Furthermore, we assume that the particles are being moved along the channel by a drag force. The natural choice for spherical particles would be a Stokes-like drag force, which in turn applies for individual particles in the limit of very small packing fraction (when porosity $\epsilon$ tends to one). For a packed suspension, the drag force is corrected by a voidage function $f(\epsilon)$ accounting for the presence of surrounding particles. Thus, particles are pushed by a corrected Stokes drag force of the form

\begin{equation}
    \mathbf{F}_i^{\rm S} = f(\epsilon) 3 \pi \eta d ( \mathbf{v} - \mathbf{u}_i ),
\end{equation}

where $\mathbf{v}$ is the fluid velocity and  $\mathbf{u}_i$ is velocity of particle $i$. Following \citet{vanderhoef2005lattice}, we use a voidage function $f(\epsilon) = 10 \frac{(1-\epsilon)}{\epsilon^{2}} + \epsilon^2 \left( 1 + \frac{3}{2} \sqrt{1 - \epsilon } \right)$, since it has been proven to model the drag force over a large range of porosities, in particular for dense suspensions.

\begin{figure*}
\includegraphics[width=\textwidth]{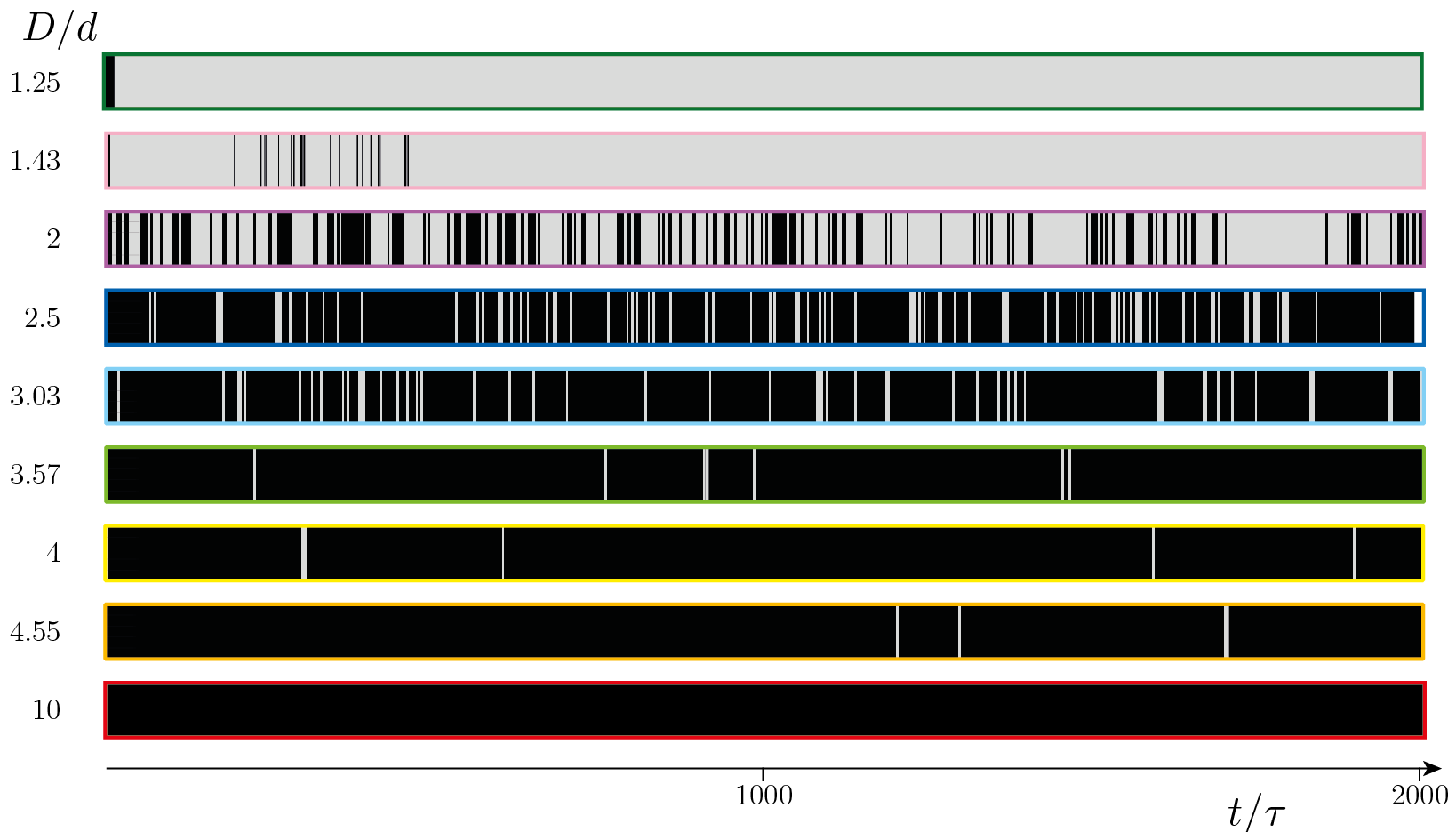}
\caption{Spatiotemporal diagrams at the constriction neck for various $D/d$. Grey : clog, no particle flowing through the constriction neck; Black : particles flowing through the neck. Each black vertical line represents the time when a particle has escaped the constriction. 
The width of every diagram line is given by the Stokes time of the particle, i.e., the time it takes a particle to move a distance equal to its own diameter at an average velocity of $\bar{v} = 5$~mm/s.}
\label{fig:3}
\end{figure*}

To account for spatial variation of porosity within the channel geometry, the porosity $\epsilon$ is estimated by dividing the channel into three parts and computing the so called local stripped porosity: (1) at the beginning of the channel, (2) in the middle of the channel and (3) close to the constriction. 
Dividing the channel in a larger number of sections, or computing a local porosity using a coarse graining approach both ended in negligible variations of the porosity once the suspension reaches its packed state. Therefore, the former, simpler and faster method was chosen.

Flow and particle properties are chosen to match the experiments: for the polystyrene particles we employ a Young modulus and Poisson ratio of $E = 3 \: \mathrm{G Pa}$ and $\nu = 0.35$ respectively. Furthermore, a diameter dependent particle roughness of $\xi = 0.005 d$ and the macroscopic sliding, rolling and torsion friction coefficients were assumed to be identical and equal to 0.6, 
 which is in the range of the typical sliding friction coefficient for frictional particles \cite{mari2014shear, gallier2014rheology, chialvo2012bridging, clavaud2017revealing}. 
 In the experiments, liquid and particle have a matching density of $\rho = 1062 \: \mathrm{kg/m^3}$ to avoid buoyancy effects, and the liquid viscosity is $\eta = 1.8 \times 10^{-3} \: \mathrm{Pa \: s}$, which is also included in the numerical model. More details about the experimental methods can be found in Appendix \ref{app1}.

The simulations consist of a flowing suspension of monodisperse particles of varying diameter $d = 10$, 22, 25, 28, 33, 40, 50, 70 and $80 \: \mathrm{\mu m}$. Adopting a neck height of $D = 100 \: \mathrm{\mu m}$ this corresponds to neck-to-particle ratios $D/d = 10$, 4.55, 4, 3.57, 3.03, 2.5, 2, 1.43 and 1.25 respectively. In the case of particles of $d = 33 \: \mathrm{\mu m}$ ($D/d$=3.03), showing the characteristic intermittency regime, the packing fraction typically reaches values of approximately 0.5, with circa 1000 particles simultaneously simulated in the channel. 
On the other end, for the largest particle case of $d= 80~\mathrm{\mu m}$, the average packing fraction was approximately 0.4, with around 60 particles simultaneously in the channel. Note that the choice of neck-to-particle ratios is chosen to match experimental data in the literature \cite{Marin2018,Souzy2020,souzy2022role}, but also involves a range where the system is clearly three-dimensional ($D/d \gtrsim$ 3) while another one is quasi-two-dimensional ($D/d<3$).
The time step value is particle size dependent as aforementioned, but always remains of the order of $10^{-6}$ s. Simulations are run over typically one million time steps and repeated tens of times with different random seeds corresponding to various initial positions of the injected particles. These are inserted at the left boundary of the channel at a constant rate to keep the maximum packing fraction attainable. Notice that when the channel is full, we make sure to not overlap particles during insertion. The particles are erased once they escape the constriction.


\section{Results}\label{sec:numericalresults}

Figure \ref{fig:2} presents successive snapshots of a typical experiment, for $D/d = 3.03$. In particular, the flow has become interrupted by the spontaneous formation of arches spanning the bottleneck (top panel in Fig. \ref{fig:2}). At this moment, particles are still experiencing a drag force, perturbing the arches which may eventually collapse. If this happens, the flow of particles is resumed (middle panel) and a burst of particles flowing through the constriction develops until a new clog arrests the flow again (bottom panel). The color coding shows in dark blue immobile particles, in lighter blue slow and in red fast moving particles.

The overall intermittent behavior, in which several flowing and arrested periods of time alternate, can be better visualized in the spatio-temporal diagrams in Figure \ref{fig:3}. There, in order to analyze the different regimes of particle flow, we report the results obtained over the range of neck-to-particle ratios $10\geq D/d\geq 1.25  $. Spatio-temporal diagrams are constructed to visually represent the passage of successive particles through the neck constriction: a black vertical line represent a particle escaping through the constriction, while grey regions are particle flow interruptions when no particle escapes, i.e., clogs. The thickness of the vertical black lines width is given by one Stokes time $\tau$, the time a particle takes to travel its own diameter length. Such a diagram is the numerical version of the spatio-temporal diagrams discussed by \citet{zuriguel2017clogging}, and constructed experimentally in \citet{Souzy2020} and \citet{Zuriguel2014}. 
The diagrams clearly reveal a qualitative difference in the flow behavior: from a clogged situation for larger particles (top diagram) to uninterrupted particle flow for the smallest ones (bottom diagram). For $D/d=1.25$ only few particles escape before the particle flow rate is interrupted (in grey) and a permanent clog is formed which last until the end of the simulation. For $D/d=1.43$, following few short intermittent bursts where particles escape in small numbers (in black), a permanent clog is eventually formed. For $D/d\geq2$ particles continuously keep on escaping in bursts. As $D/d$ is increased, the flowing intervals become longer and more abundant. This regime persists until the particle flow becomes continuous for $D/d\geq 10$, so the bursts intermittency becomes immeasurable. At this point, a minimum clog time $T_{\rm{min}}$ needs to be defined in order to discern clogging from flowing: a flow interruption longer than $T_{\rm{min}}$ separates the end of a burst and the beginning of another. Note that given the discrete nature of the system, defining an arrest time threshold to set apart successive bursts is not straightforward. This is done by looking at the distribution of times $T$ between the passage of two consecutive particles. From those distributions, and using the Clauset-Shalizi-Newman method \cite{Clauset2009}, the characteristic minimum time $T_{\rm{min}}=\tau/2$ is obtained, which corresponds to the time it takes for a particle to travel its own radius (more details in Appendix \ref{app2}). For $D/d\geq 10$, no time lapse $T$ between the passage of consecutive particles is reported to be larger than $T_{\rm{min}}$.

\begin{figure}
\includegraphics[width=0.5\textwidth]{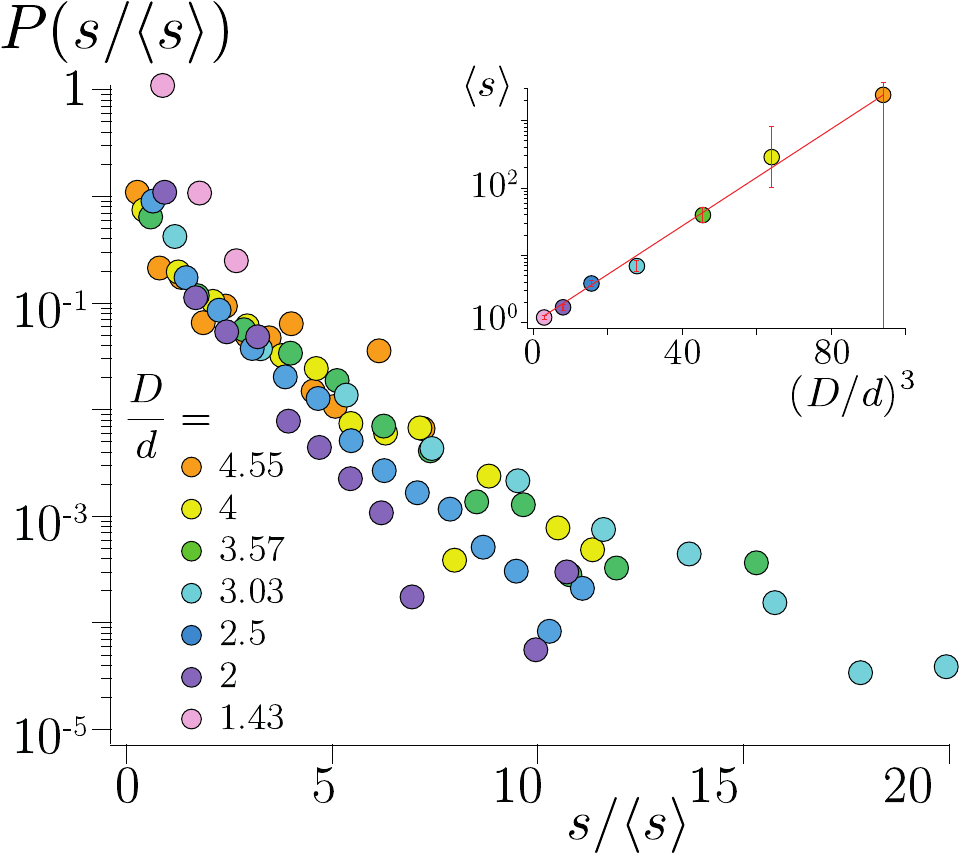}
\caption{Distribution of the normalized number of particles escaping per burst $s/\langle s \rangle$, which follow a similar exponential distribution over the range of investigated $D/d$. Inset: the average $\langle s\rangle$ plotted as a function of $(D/d)^3$, with the error bars standing for the standard deviation. The red line corresponds to the best exponential fit, signature of the predicted exponential trend expected by \citet{Thomas2015} for a transition from a regime where clogs occur very frequently (only few escapees before a new clog develops for $D/d=1.43$) to a regime of almost continuous particle flowing (thousands of particles escape before a clog develops for $D/d=4.55$).}
\label{fig:4}
\end{figure}

To quantify the intermittent dynamics in what follows, we will analyse separately the
arch formation and destruction processes by looking at the statistics of burst sizes and
arrest times, respectively. These two measurable quantities are proxies of these two crucial mechanisms: the number of escapees per burst is an indicator of the probability of clogging, while the arrest time characterizes the lifetime of a clog once it is formed, thus being an indicator of the unclogging probability.

\begin{figure*}
\includegraphics[width=\textwidth]{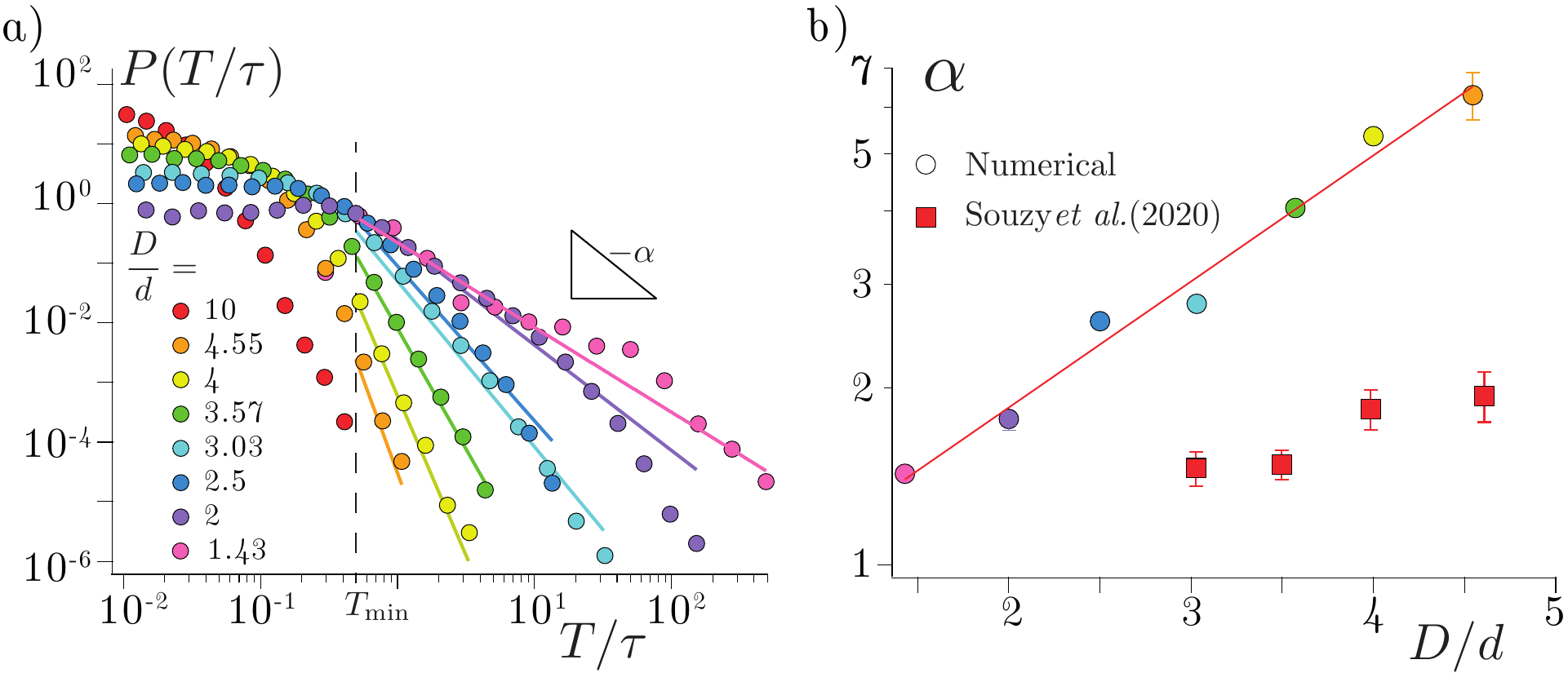}
\caption{a) Distribution of the arrest time lapses $T$ normalized by the Stokes time $\tau$. The lines correspond to the best power-law fits with their exponent $\alpha$, as determined using the Clauset-Shalizi-Newman method \cite{Clauset2009}. b) $\alpha$ as a function of the neck-to-particle size ratio D/d. The exponent can only be defined when the flow is intermittent and comparison with previous experimental measurements in a similar configuration reveal quantitative value mismatch yet similar trend with increasing $D/d$. The red line corresponds to an exponential fit $\alpha\propto e^{0.5 D/d}$.}
\label{fig:5}
\end{figure*}

\emph{Arch formation.} Similarly to pedestrians \citep{Garcimartin2016pedestrians}, animal flocks \citep{Garcimartin2015} and avalanches  \citep{Fisher1998}, the number of entities escaping per burst has been experimentally found to follow an exponential distribution in constricted flow of suspensions \citep{Souzy2020, souzy2022role}. While monitoring the number of particles per burst is not an easy task experimentally given that particles overlap frequently in this three dimensional configuration, achieving such quantitative measurement using numerical simulations is trivial as the position of each single particle is monitored at each time. The corresponding distribution $P(s/\langle s\rangle)$ of the number of escapees per burst normalized by the average number of escapees, which are shown on figure \ref{fig:4} for various $D/d$, are consistent with the previous experimental measurements reported in the literature. Such exponential distributions of $s/\langle s\rangle$ reveal that the arch formation process is Poissonian and consequently it can be described using simple stochastic models, considering that an arch develops when a sufficient number of randomly arriving particles reach the constriction in the appropriate arrangement \cite{zuriguel2003jamming,Marin2018}. Similarly, Thomas \& Durian \cite{Thomas2015} proposed a model for dry systems in which they showed that the discharged particle mass grows as an exponential function of hole diameter (to the power of the dimensionality of the system), which is also in agreement with the results reported in the inset of figure \ref{fig:4}, highlighting that the average number of escapees per burst is well approximated using an exponential fit $\propto e^{(\frac{D}{d})^3}$.  Both models are actually compatible with the idea that new \emph{microstates} \cite{Thomas2015} in the vicinity of the constriction are continuously and randomly sampled while particles arrive, until a stable arch is eventually found. A first consequence of such clogging mechanism contemplated by both models is that there is no sharp clogging transition for a given critical outlet size in the sense that there is always a non-zero probability for a clog to occur. A second consequence is that the average number $\langle s\rangle$ of escaping particles per burst is therefore a good proxy for the probability of clogging: the higher $\langle s \rangle$, the smaller the probability of arch formation.

A closer look at the distribution $P(s/\langle s\rangle)$ shows a substantially higher probability for the smallest particle bursts (typically bursts of 1 to 3 particles). Interestingly, such feature has already been reported experimentally in three-dimensional flow of constricted suspension \cite{Souzy2020} and in three-dimensional silos \cite{zuriguel2005jamming} and shows that, relatively often, a small number of particles may manage to escape through the arch without destabilizing it.  We interpret this feature as a result from the system's inherent three-dimensionality, as such behavior is not reported for two-dimensional configurations. It is rather remarkable to notice that although using an unresolved method for modelling the viscous drag, simulations are still able to capture such fine details which are experimentally observed.


\emph{Arch destruction.} To investigate the unclogging process, we now analyse the probability distributions of time lapses $T$ between the passage of consecutive particles. Such an approach has been extensively implemented in previous studies on intermittently flowing systems, such as pedestrian crowds \citep{Helbing2005, Krausz2012,Garcimartin2016pedestrians}, hungry sheep herds \citep{Garcimartin2015}, mice escaping a water pool \citep{Saloma2003}, or vibrated silos of dry granular material \citep{Janda2009, Lastakowski2015}. Interestingly, in such systems the distribution of arrested time lapses exhibits a power-law tail $P(T)\propto T^{-\alpha}$, a signature of systems susceptible of clogging \citep{Zuriguel2014,Zuriguel2020}. Furthermore, the value of the exponent $\alpha$ can be directly related to the long-term behaviour of the system: the average time lapses $\langle T\rangle$ can only be defined for distributions fulfilling $\alpha>2$, while $\langle T\rangle$ diverges for $\alpha\leq2$. This feature has therefore been interpreted as a transition to a scenario in which a permanent clog will eventually develop.  For $\alpha\leq2$, there is a non-zero probability of observing everlasting clogs, while for $\alpha>2$, the system can be temporary blocked due to the formation of clogs but no arch will persist infinitely. More detailed discussions can be found in \citet{Zuriguel2014}, \citet{Zuriguel2020}, or in \citet{Garcimartin2021}.

Figure \ref{fig:5}a presents the probability distribution of the arrest lapses obtained for various $D/d$ from the spatio-temporal diagrams shown in Figure \ref{fig:3}. The distribution $P(T/\tau)$ exhibits the characteristic power-law tail $P(T/\tau)\propto(T/\tau)^{-\alpha}$. Note that we have been able to measure time lapses up to two orders of magnitude larger than the Stokes time, and that for each fixed value of $D/d\geq 2$ more than $\sim 1000$ bursts have been analysed. For $D/d=1.43$, as the flow is composed of few bursts of escaping particles before a permanent clog develops within a simulation, statistics are restricted to $\sim 150$ bursts.
Finding the right parameters for power-law tails can easily suffer from arbitrary biases, therefore the exponent $\alpha$ of the power-law tail is obtained using the rigorous and widely accepted Clauset-Shalizi-Newman method \citep{Clauset2009}, which also yields the minimum time lapse $T_{\rm{min}}$ from which the power-law fit is valid.
As mentioned earlier, note that for $D/d=10$ no time-lapse $T\geq T_{\rm{min}}$ is reported, signature of the transition to continuous particle flow for large neck-to-particle aspect ratio.

Figure \ref{fig:5}b presents the value of $\alpha$, for various neck-to-particle size ratios $D/d$. The error bars in the vertical axis follow directly from the Clauset-Shalizi-Newman method \citep{Clauset2009}, and they represent the uncertainty in the estimate fit to a power law. Note that for larger $D/d$, the fit is performed over scarcer events as most of the time lapses between the passage of consecutive particles lay below $T_{\rm{min}}$, thus resulting in larger uncertainty in the $\alpha$ value. 

The first important thing to notice is that,  as previously reported \citep{Souzy2020}, the value of $\alpha$ is remarkably sensitive and increases significantly to the neck-to-particle size ratio. This is expected: the smaller the particles, the higher the value of $\alpha$, thus the higher is the probability of short-lived clogs. This highlights the fact that arches composed of more particles (large $D/d$) are less stable, and thus more prone to break due to the perturbations induced by the interstitial flow. In other words, shorter arches are stronger than longer ones.
Remarkably, over the explored range of $D/d$ for which an exponent $\alpha$ could be estimated, values of $\alpha>2$ were found, thus indicating that the intermittent regimes would continue indefinitely, with a zero probability of permanent clogs which would persist endlessly. This is similar to other scenarios where clogging transitions have been reported based on the power-law tails of the arrest times, like vibrated silos \citep{Janda2009}, Brownian particles \citep{Hidalgo2018}, pedestrians \citep{Zuriguel2014}, and self-propelled robots \citep{Patterson}. All those systems reported a fairly smooth transition from an intermittent clogged state ($\alpha\leq2$) to a continuous flow, passing through a region of intermittent flow with $\alpha>2$. Interestingly, the values of $\alpha$ from the simulations are well fitted by an {exponential fit $\alpha\propto e^{0.5 D/d}$} in the explored range of $D/d$, highlighting again how crucial the neck-to-particle size ratio parameter is both to the arch formation \emph{and} destruction processes.

The second thing to notice is that there is a significant quantitative mismatch between the values of $\alpha$ reported for the numerical simulations with those reported for the experiments of \citet{Souzy2020}. In this respect, note that  \citet{souzy2022role} reported that the value of $\alpha$ is independent of the imposed flow rate in volume controlled configuration. Therefore, when comparing the results of the simulations to the experimental results obtained under constant flow rate configuration, the quantitative mismatch cannot be attributed to a mismatch between the flow rate imposed within the experiments and the numerical simulations. The higher values of $\alpha$ in the numerical results imply a higher probability of arch destruction, thus a lower arch stability when compared to experiments. We could speculate about different reasons for such mismatch. One reason could be attributed to the porosity estimation $\epsilon$, or to the voidage function $f(\epsilon)$, which may overestimate the Stokes drag force, thus leading to an enhancement of the arch destabilization mechanism. Another hypothesis could be found in the presence of ``fast channels'' \cite{Siena2019Porous} within the dynamic porous network in experiments, which cannot be reproduced in the current simulations due to the lack of coupling between the fluid flow and the particle network.


Another cause for this discrepancy may also lie on how the friction coefficient is accounted for in the simulations. Three distinct friction coefficients can be set: sliding, rolling and torsion. As aforementioned  these parameters were all set to an identical value of 0.6. However, one could discard such assumption and argue that one of the specific friction coefficient should be tuned to a greater value than the others. Increasing the value of the rolling coefficient friction for instance, should make it harder for particles to roll on each other and stabilize arches. 


\section{Conclusion}\label{sec:conclusion} 

Performing numerical simulations of the intermittent flow of particulate suspensions through a constriction using an advanced discrete particle solver and an approximate numerical model for the liquid drag, our simulations yield qualitatively good results when compared to experimental investigations. An intermittent flow of successive bursts of released particles is reported for $1.43\leq D/d\leq 4.55$. For $D/d>4.55$, the flow of particle is found to be continuous with no discernible particle flow interruption. Surprisingly, the intermittent flow regime is also reported down to $D/d=1.43$, signature that the particle flow is enhanced in the simulations compared to the experiments. Consistent with the currently accepted understanding of the arch formation process \cite{zuriguel2003jamming,Thomas2015}, the average number of escapees per burst follows a Poisson distribution with the neck-to-particle size ratio, i.e., $\langle s\rangle$ follows an exponential increase with $(D/d)^3$. 
However, the quantitative values of $\langle s\rangle$ lie below what is experimentally reported \cite{souzy2022role}, suggesting that the \emph{arch construction} process is somehow more likely in the simulations. Yet, subtle features such as a higher probability for the smallest particle bursts (already reported in three-dimensional flow of constricted suspensions \cite{Souzy2020} but also in granular silos \cite{zuriguel2005jamming}) are remarkably captured within the simulation, highlighting the striking phenomenological agreement.
Regarding the \emph{arch destruction} processes, we report larger values of the exponent $\alpha$ from the simulations compared to the experiments, a direct signature of shorter-living arches. Consequently, both the arch construction and the arch destruction are enhanced \emph{in silico} when compared to the experiments. 

Understanding the fundamental reasons for such puzzling discrepancies is yet to be investigated and require additional study. The current results reveal that resolving the fluid flow is not necessary to mimic the rich phenomenology of particulate suspensions flowing through a constriction, offering a promising cost-efficient numerical method to study the phenomenology of such a complex problem. A tool like this could be particularly useful to investigate the effect of various parameters which are experimentally very challenging to explore on the overall flow behaviour, such as the effect of particle roughness, the constriction angle, the particle softness, deciphering the respective importance of hydrodynamic forces, or extensively exploring the intermittent flow transition by varying the neck-to-particle size ratio. However, our results indicate that resolving the interparticle flow with a numerical CFD solver may be necessary when a direct comparison with experiments is required.



\section{Acknowledgements}
This work was supported by the ERC (European Research Council), Starting Grant (grant agreement No.678573). The authors would like to acknowledge the motivation for this work and insights from Iker Zuriguel and Raul Hidalgo. The authors would also like to acknowledge fruitful discussions with Prof. Stefan Luding and his continuous support.

\appendix\section{ Experimental methods}\label{app1}

The experimental results used as a benchmark for the numerical results are extracted from \citet{Souzy2020} and \citet{souzy2022role}. The experimental set-up, also used in \citet{Marin2018}, consists of a  single transparent straight channel of borosilicate glass (isotropic wet etching, Micronit microfluidics) with a rectangular cross-section of $100 \times 400\,\SImum^2$ which reduces to an almost square cross-section of $100\times110\,\SImum^2$ to form the neck. A linear narrowing of the channel with a half-angle of 60\degree achieves the constriction, such that the fluidic system forms a two-dimensional nozzle converging towards the neck. Particles and liquid have been chosen to avoid buoyancy effects, particle aggregation and particle deposition at the micro-channel walls: the suspension consists of monodisperse spherical polystyrene particles (Microparticles GmbH) of diameter $d$ which is varied from 19.0 to 41.1 $\SImum(\pm3\%$). Particles are stabilized with negatively charged sulfate groups in a density-matched 26.3 wt\% aqueous solution of glycerine, with a density $\rho=1062\,$kg/m$^3$ 
\citep{Volk2018density}. The charged sulfate groups confer them a small negative surface potential (on the order of $-50$ mV) but sufficient to prevent both their agglomeration and their adhesion to the channel walls. The suspension is prepared with a particle volume fraction of about $2\%$, then inserted in the device and driven downstream the constriction towards a filter which only allows the fluid to flow through. Particles are therefore initially concentrated in that position.

An experiment starts when the flow is reversed and particles are dragged by the fluid towards the constriction. Note that although the experimental system is designed to flow either in pressure or volume-controlled driving, the comparisons were done with experimental results obtained in an imposed volume-rate configuration.
Particles flow towards the constriction forming a compact and long ``column'' of particles, and the suspension is imaged with a high-speed CMOS camera (PCO.dimax CS1) coupled to an inverted microscope (Nikon Instruments, Eclipse TE2000-U). 

Note that the reason for using such particle size range (between 10 and 50 $\SImum$) is dual: on the one hand, we avoid colloidal particle interactions and Brownian motion. On the other hand, increasing the particle size further would also involve handling larger volumes of fluid, i.e., larger Reynolds number $Re$ and higher working pressures. Therefore, the range of particle size chosen allows to work with highly monodisperse particles interacting mainly by hydrodynamic interactions and low-pressure solid contacts, manipulated via microfluidic technology, which allows to obtain a high degree of control and reproducibility difficult to achieve experimentally at other length scales. Each experiment was typically repeated $\sim30-50$ times, where each recorded run typically monitors tens of clog formation/destruction events.

\section{Minimum arrest time }\label{app2}
Given the discrete nature of the system, defining an arrest/clog time threshold to set apart successive bursts is not straightforward. This is done by looking at the distributions $P(T)$ of times between the passage of two consecutive particles, which exhibit power-law tails. Finding the right parametric fit for power-law tails can easily suffer from arbitrary biases when trying to characterize experimental data sets. To tackle such issue, the exponents $\alpha$ of the power-law tails are obtained using the rigorous and widely accepted Clauset-Shalizi-Newman method \citep{Clauset2009}. This method also yields the estimated error of the fit, which is used to set the error bars on Fig \ref{fig:5}b. The method can also either be used leaving the minimum time lapse $T_{\rm{min}}$ from which the power-law fit is performed as a free parameter which will be determined by the method, either get the value of $T_{\rm{min}}$ directly provided as an input.
When using the Clauset-Shalizi-Newman method over the data leaving $T_{\rm{min}}$ as a free parameter, we find that $T_{\rm{min}}\approx 0.5 \tau\pm 0.1\tau$ over a wide range of $2\leq D/d\leq 4.55$. A value of $T_{\rm{min}}= 0.5 \tau$, which corresponds to the time it takes for a particle to travel its own radius, was therefore subsequently used as an input for the Clauset-Shalizi-method to determine the values of $\alpha$. More importantly, this value was used to define the minimum time lapse to set apart consecutive bursts: a clog event is defined as any event where the time lapse $T$ between the passage of two consecutive particles is such that $T\geq T_{\rm{min}}$.

\bibliographystyle{apsrev4-1}
\bibliography{Bibliography}

\end{document}